\documentclass[twocolumn,tighten]{aastex63}
\usepackage{graphicx}
\usepackage{gensymb}
\usepackage{longtable}
\usepackage{amsmath}

\begin{document}

\title{Deep Optical Images of the Ejecta Nebula Around the Wolf-Rayet Star WR 8 (HD 62910) }

\author[0000-0003-3829-2056]{Robert A.\ Fesen}
\affil{6127 Wilder Lab, Department of Physics and Astronomy, Dartmouth
       College, Hanover, NH 03755 USA}

\author[0000-0002-7507-8115]{Daniel Patnaude}
\affil{Smithsonian Astrophysical Observatory, MS-3, 60 Garden Street, Cambridge, MA 02138 USA}

\author[0000-0003-2588-1265]{Wei-Hao Wang}
\affil{Institute of Astronomy and Astrophysics, Academia Sinica (ASIAA), No.\ 1, Roosevelt Road, Taipei 10617, Taiwan, R.O.C.}

\author[0000-0003-3667-574X]{You-Hua Chu}
\affiliation{Department of Physics, National Sun Yet-Sen University, No.\ 70, Lienhai Rd., Kaohsiung 804201, Taiwan, R.O.C.}


\author{Jason Sun}
\affiliation{Department of Physics, National Tsing Hua University, No.\ 101, Section 2, Kuang Fu Road, Hsinchu 30013, Taiwan, R.O.C.} 



\author{Manuel C.\ Peitsch}
\affiliation{Roboscopes at e-EyE Entre Encinas y Estrellas Observatory, 
Camino de los Molinos 06340 Fregenal de la Sierra, Spain}

\author{Martin Pugh}
\affil{Observatorio El Sauce, 
Rio Hurtado, Coquimbo, Chile}

\author{Scott Garrod}
\affil{Observatorio El Sauce, Rio Hurtado, Coquimbo, Chile}

\author{Michael Selby}
\affil{Observatorio El Sauce, Rio Hurtado, Coquimbo, Chile}

\author{Alex Woronow}
\affil{Department of Geosciences,
University of Houston, 4800 Calhoun Rd. Houston,  TX 77004 USA}

\begin{abstract}

\noindent
We report the results of deep 
H$\alpha$ and [\ion{O}{3}] images
of the bright WN6/WC4 Wolf-Rayet star WR~8 (HD~62910).
These data show considerably more surrounding nebulosity than seen in prior imaging.
The brighter portions of the nebula span 
$\simeq6'$ in diameter and exhibit considerable fine-scale structure including numerous emission clumps and bright head-tail like features presumably due to
the effects of the WR star's stellar winds.
Due to the overlap of a 
relatively bright band of unrelated foreground diffuse interstellar H$\alpha$ emission, 
WR~8's nebula is best viewed via its [\ion{O}{3}] emission. A faint $9' \times 13'$ diffuse outer nebulosity is detected surrounding the nebula's main ring
of emission. Comparison of the
nebula's optical structure with that seen in WISE 22 $\mu$m
data shows a similarly clumpy structure but within a better defined emission shell
of thermal continuum
from dust. The infrared shell is coincident with the nebula's southern [\ion{O}{3}] emissions but  is mainly seen 
in the fainter outer portions of the northern
[\ion{O}{3}] emission clumps.
It is this greater radial
distance of dust emission in the nebula's northern areas
that leads to a striking off-center position of the
WR star from the IR shell.

\end{abstract}


\bigskip


\section{Introduction}

Wolf-Rayet (WR) stars are thought to be 
the helium-burning descendants of a galaxy's most massive stars with initial masses $\geq30~$M$_{\odot}$ 
\citep{Maeder1994, Conti2000, Crowther2008}.
Identified by their broad optical line emissions,
these stars have stellar winds characterized by terminal
velocities ranging from 1000 to 4000 km s$^{-1}$ and due to
their very high luminosities undergo high  mass-loss rates $\sim$ $10^{-5}$ M$_{\odot}$ yr$^{-1}$ 
\citep{Hamann2019,Sander2022}.  
During its WR phase, a massive star can release up to
$10^{51}$ erg, which is comparable to that of a supernova (SN) explosion. Consequently, WR stars can have significant effects on their local environments \citep{Abbott82, Oey1999}.

WR stars are categorized into two main subtypes based on the types of optical line emissions they exhibit;
WN stars show emission lines from helium and nitrogen ions, while WC stars show carbon and oxygen lines in addition to helium lines. Emission lines of hydrogen are
notably absent in all but a few WR stars, as are strong absorption lines \citep{Abbott1987}.
WR subtypes are also subdivided into an excitation/ionization
sequence where the highest to lowest excitation spectral features
are denoted by low to high subclass numbers. 
Thus, WN stars are arranged WN3 to WN9 based on the
relative strengths of \ion{N}{3}, \ion{N}{4}, and
\ion{N}{5} lines. Similarly, WC stars have subtypes 
based on excitation/ionization differences, ranging from
WC3 to WC9. 

A few WR stars exhibit both WN and WC like spectra in that they show strong emission lines
of carbon and nitrogen simultaneously 
\citep{Conti_Massey1989,Crowther2007}.
It is unclear if such stars are single stars in the transition between WN and WC, or if some are cases of WN + WC binaries.
Regardless, such stars are relatively rare, consistent with an expected short $\sim10^{4}$ yr transition
time-scale; see  discussion  in \citet{Hillier2021}.

Due to their strong stellar winds and high mass loss rates, 
WR stars are often surrounded by optical emission arcs or nebula `rings' \citep{Chu1981, Chu1983}. Since the first Galactic WR nebulae 
were discovered in the 1960's \citep{Johnson1965}, dozens more have
been identified \citep{Chu1991,Chu1983}. WR associated nebulae are categorized into three groups based on their formation mechanisms. These are:
W-type for a wind-blown ISM bubble, E-type for stellar mass ejection nebulae, and  R-type for a radiatively excited H~II region.
Most WR nebula are associated with WN type stars, and WN8 stars in particular 
\citep{Chu2016}.

Of special interest are the E-type ejecta WR star ring nebulae due to their
formation by means of substantial and violent mass loss episodes in the recent history of the star's mass loss evolution. Unlike wind-blown (W) or H~II region (R) types of WR ring nebulae, E-type ejecta nebula tend to
more clumped and irregular in morphology
\citep{Chu1981, Chu1991}.
These nebulae also show greater nitrogen and helium abundances relative to the ISM consistent with CNO cycle processed  material \citep{Kwitter1984, Esteban2016}.

The eighth WR star listed in the well referenced
\citet{vanderHucht2001} catalog of Wolf-Rayet stars
is the WR star HD~62910 (V = 10.1 mag), hence giving it the WR~8 identification.
Located in the southern hemisphere constellation of Puppis, 
it lies at 
an estimated distance of
$3.52\pm0.16$ kpc \citep{Bailer2021}.
It has been classified as a WN6/WC4 due to its spectrum showing both WN and WC spectral features \citep{vander1981,Willis1990}.

Some WN/WC stars,
such as WR 145 and WR 153, are known to be spectroscopic binaries, and there are indications that WR~8 might also be a binary. 
While \citet{Niemela1991} and \citet{Marchenko1998}
found strong evidence for WR~8's binary nature,
\citet{Willis1990} 
found a good correlation of line widths and excitation
potential for both WN and WC spectral features suggesting they formed in a single stellar wind.
Although a single star conclusion was supported by \citet{Deshmukh2024}, their instrument's resolution
may not have been sufficient to rule out the binary nature of WR~8 and \citet{Sander2012} was unable to fit its spectrum using a single-star model.


Despite several searches looking for
a nebula associated with WR~8
(e.g., \citealt{Heckathorn1982, Chu1983, Marston1994a}),
a faint and clumpy nebulosity was only recently discovered by 
\citet{Stock2010} (see Fig.\ 1) examining the 1.2 m Schmidt AAS/UKST SuperCOSMOS 
H$\alpha$ survey images of the southern galactic plane
also known as the Southern H$\alpha$ Survey or SHS
\citep{Parker2005}. Due to the nebula's highly clumpy appearance, 
the WR~8 nebula was
classified as an E or ejecta type WR ring nebula 
\citep{Stock2010, Stock2011}, thus a member  of the rarer type
of WR nebulae.

A nebula around WR~8 was independently
detected by \citet{Wachter2011} in the infrared
from an inspection of WISE 22 $\mu$m images.
Subsequent optical and infrared spectra showed
enhanced nitrogen abundance 
\citep{Stock2011,Stock2014} characteristic
of material lost following the star's red supergiant (RSG) 
or luminous blue variable (LBV) phase 
and consistent with its E-type ejecta WR subtype.
Infrared 22 $\mu$m WISE images of WR~8's nebula shows
a nearly complete ring of emission 5.3$'$ in diameter but one that  is centered almost an arcminute north of the WR star 
\citep{Toala2015}.

Here we report the results of deep 
H$\alpha$ and [\ion{O}{3}] images
of the Wolf-Rayet star WR~8
which reveal the nebula in far greater detail
than seen in prior published images. These images 
reveal a complex nebula composed of emission clumps 
and radially aligned streamers plus a faint outer 
emission halo in [\ion{O}{3}] emission.
Our observations are described in
$\S2$ with results and discussion
presented in $\S3$.

\begin{figure}[]
\centerline{\includegraphics[angle=0,width=9.0cm]{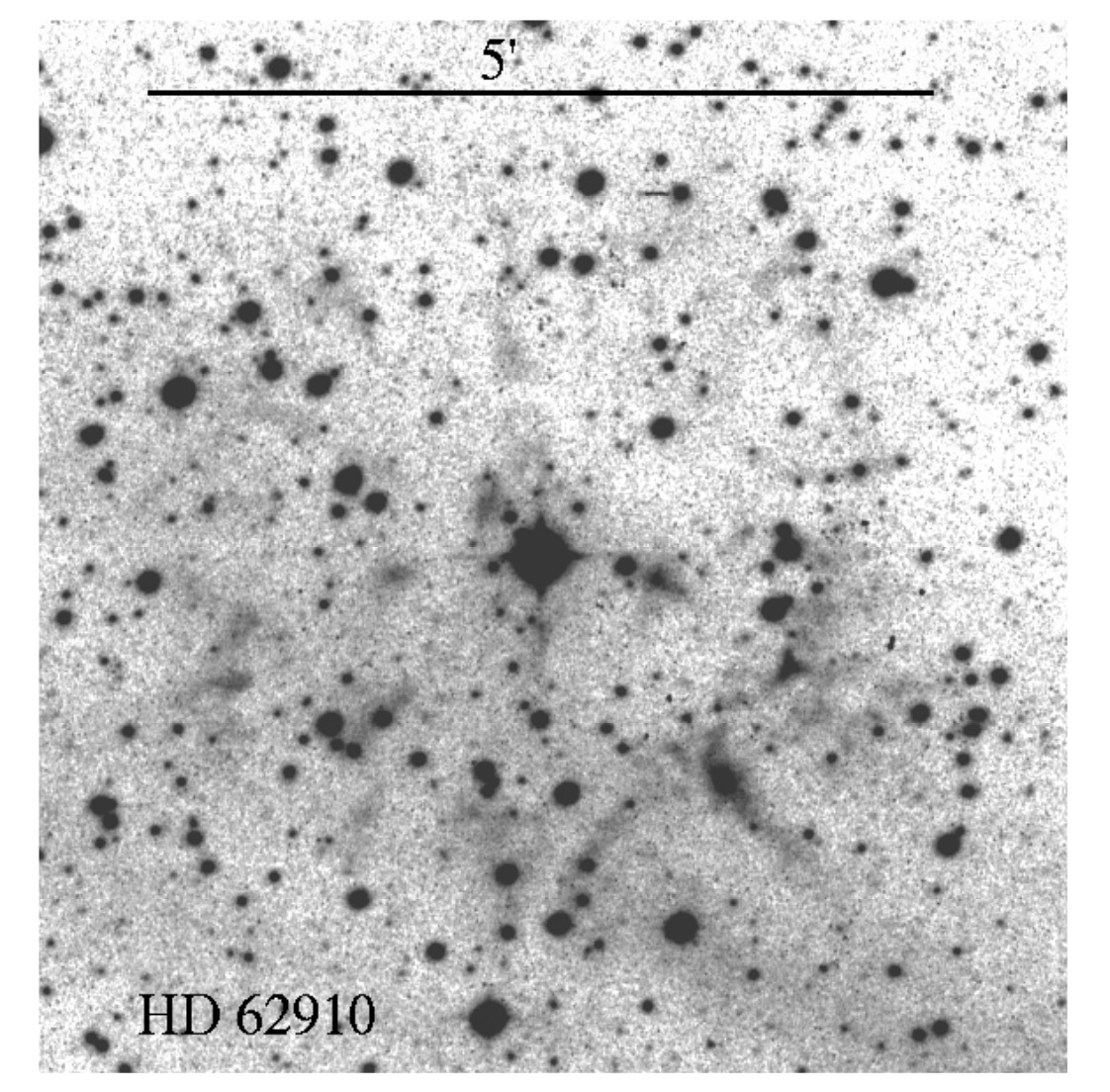}}
\caption{Discovery H$\alpha$ image showing faint nebulosity around and in front of WR~8  (from \citealt{Stock2010}).  \label{Stock}  }
\end{figure}

\section{Observations}

Due to the absence of any published H$\alpha$ images of WR~8  since the 2010 discovery of its associated nebula (\citealt{Stock2010}; Fig.\ 1), plus the lack of any [\ion{O}{3}] images of this WR nebula,
exploratory images of WR~8 were
obtained in December 2023
using a 0.5 m Planewave CDK20 f/6.8 telescope 
at the Yushan National Park in Chia-Yi, Taiwan. 
A series of $90 \times 480$ s exposures in both H$\alpha$ + [\ion{N}{2}] 
$\lambda\lambda$6548,6583 and 
[\ion{O}{3}] $\lambda$5007 filters were taken using 5 nm passband 
filters with total exposure times of 12 hr and 14.7 hr, respectively.
Reduction of these images detected 
considerable but very faint [\ion{O}{3}]
emission in addition to the nebula's already known H$\alpha$ emission.
These data indicated the presence of even fainter [\ion{O}{3}] emission farther out around the WR star plus
hints of radial streamlines going from the nebula's inner region out to the edge of the faint extended [\ion{O}{3}] emission.  

These observations lead to much deeper images taken 
using a 0.60 m Planewave CDK24 telescope
at  the Observatorio El Sauce located in Rio Hurtardo Valley, Chile. 
These were taken using narrower 3 nm
passband H$\alpha$ and [\ion{O}{3}] $\lambda$5007 filters 
and a Moravian C3-61000 CMOS camera 
($9576 \times 6388$ pixels)
yielding an image scale of
$0\farcs39$ pix$^{-1}$ and a $20' \times 13'$ FOV.
Images were taken over 16 photometric and non-photometric
nights in February and March 2024. 
Total exposure times in H$\alpha$ and [\ion{O}{3}] were 9.3 hr ($28 \times 1200$ s)
and 16.3 hr ($49 \times 1200$ s), respectively. Broadband
R,G,B filter images were also obtained with single
20 minute exposures for each filter. These data were reduced using
commercial imaging processing software including PixInsight and
Astropixel Processor. A subset of these images were also reprocessed separately using a different set of software.

Additional images were obtained in 
late December 2024 using a
0.7 m Planewave CDK700 telescope also at
Observatorio El Sauce in Chile. Using a
Moravian C5-100 CMOS camera 
with $11664 \times 8750$ pixels and 3 nm passband 
filters, images with
total exposure times of 14.2 hr for 
H$\alpha$ and 19.2 hr for [\ion{O}{3}] were obtained. 
These images were processed
with Photoshop and PixInsight.

Lastly, as a check on some of the finer 
features of the WR~8's nebulosity,
H$\alpha$ + [\ion{N}{2}] and [\ion{O}{3}] $\lambda$5007
filter images were obtained
in early January 2025 using the IMACS f/4.3 camera \citep{Dressler2011} on the Baade 6.5m telescope at Las Campanas, Chile. This camera consists of 
a 8k$\times$8k mosaic covering an
unvignetted field out to a radius of 12 arcmin. Three sets of 600 s exposures in each filter were taken offset from
WR~8 and dithered so as
to cover all the CCD chip gaps. 
Seeing during these observations ranged from $0.5''$ to $0.7''$. These data were processed by standard IMACS data reductions for mosaic imaging.


\section{Results and Discussion}

Our H$\alpha$ and [\ion{O}{3}] $\lambda$5007 images reveal a complex and striking  nebulosity surrounding the bright WR~8 star, one that is far more extensive than 
previously realized. The images also show
the presence of a faint halo of [\ion{O}{3}] emission outside the nebula's main emission shell.
Because of the nebula's 
extreme faintness, below 
we present the WR~8 nebula  shown 
at different contrasts and brightness stretches to highlight
various structural properties of the WR~8 nebula.

We begin with Figure 2 which shows our 0.6~m telescope H$\alpha$ and [\ion{O}{3}] images of WR~8 nebula  at low contrast so as to emphasize the similarity of the nebula's H$\alpha$ and [\ion{O}{3}] line emissions as well as the relative brightnesses of various nebula features in these emission lines. 
Even at this low intensity display, the structure of the WR star's nebula seen in our
in H$\alpha$ image is clearer and of higher resolution than that of the discovery SHS image (Fig. 1). 


\begin{figure}[t]
\begin{center}
\includegraphics[angle=0,width=8.6cm]{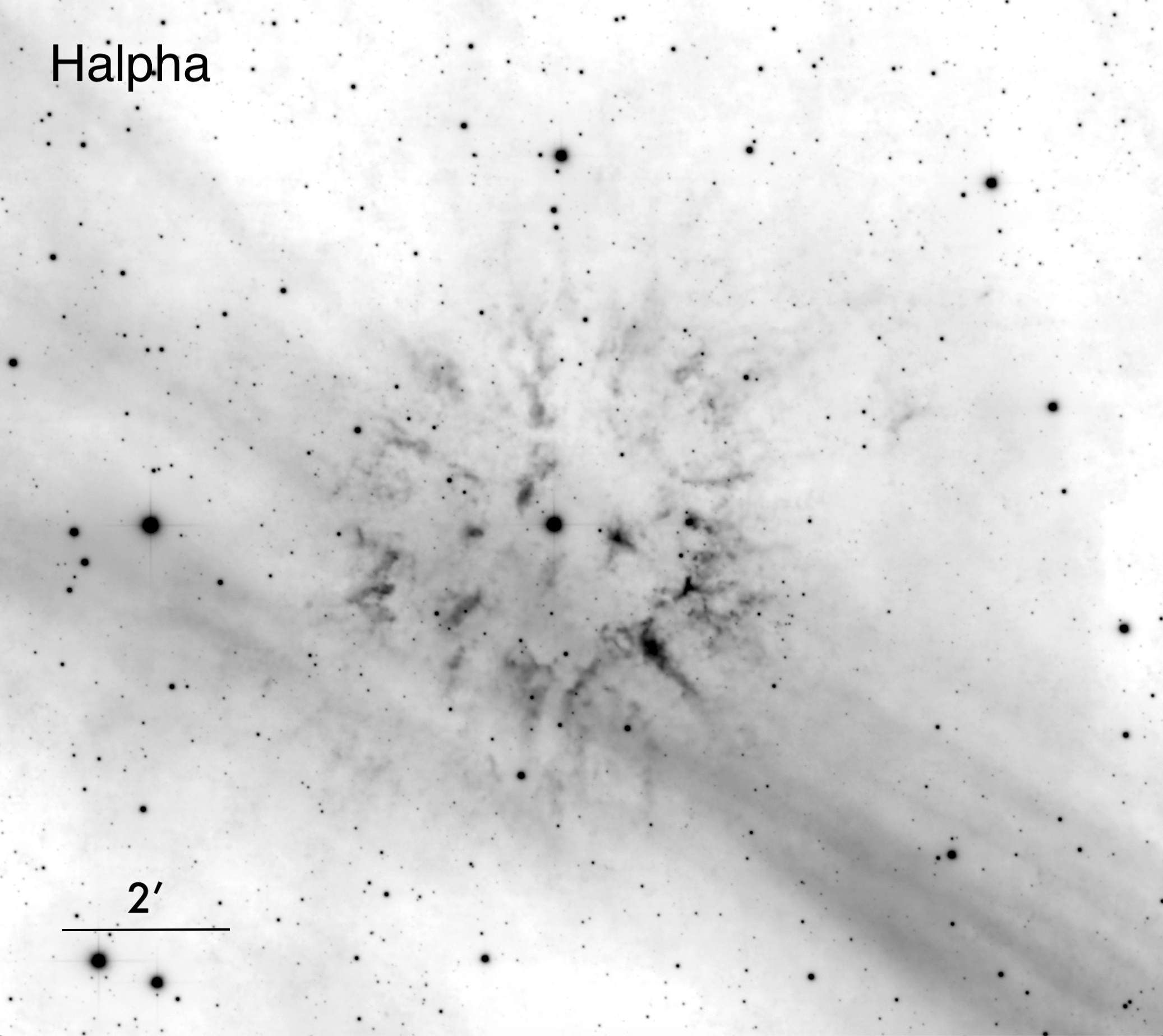} 
\includegraphics[angle=0,width=8.6cm]{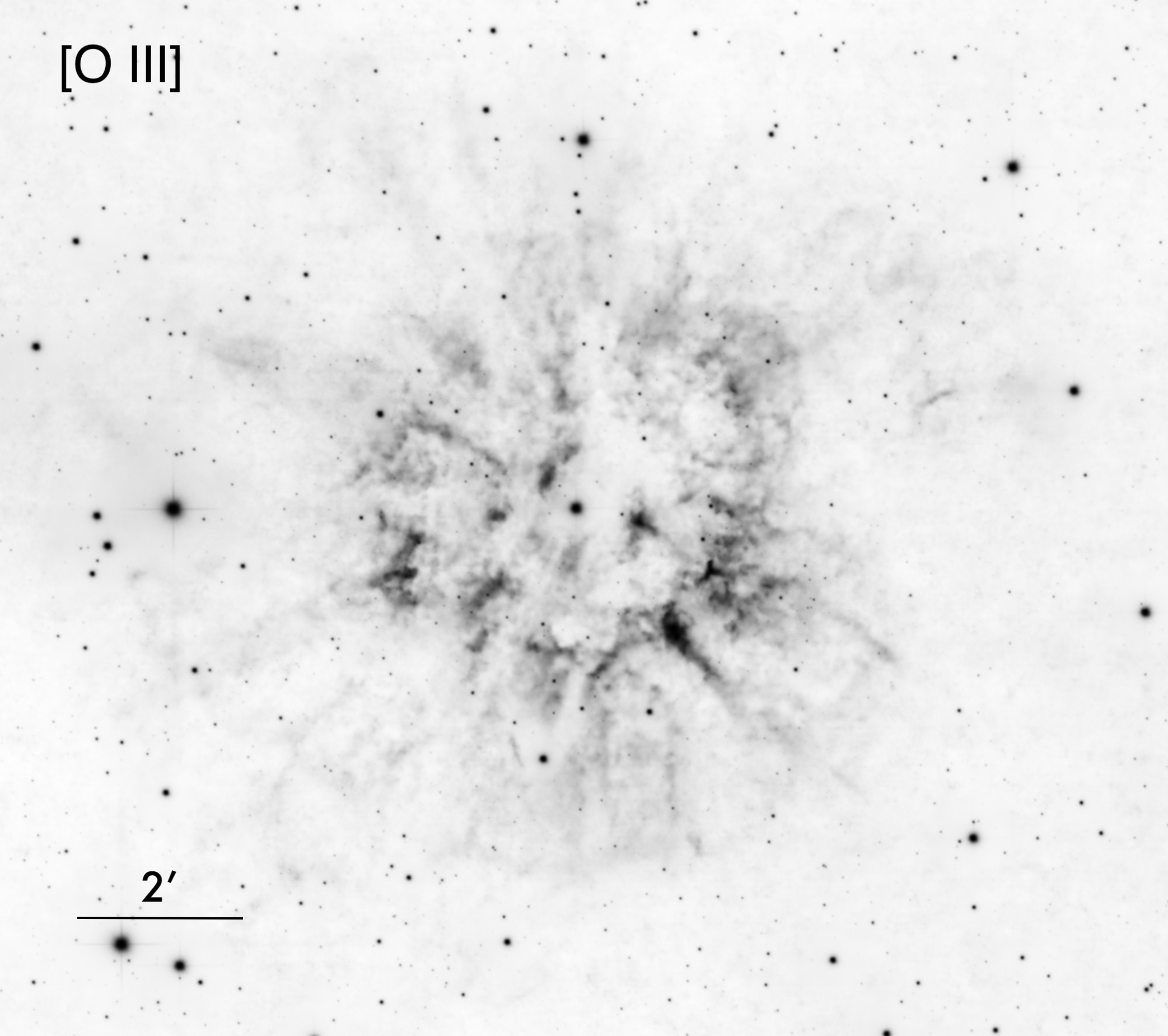}
\caption{H$\alpha$ and [\ion{O}{3}] images of the WR~8 nebula showing a comparison of the nebula's relative brightness in these line emissions. The broad band of diffuse emission seen in the H$\alpha$ image is unrelated background/foreground ISM emission. North is up, east to the left. 
\label{Fig2}
} 
\end{center}
\end{figure}

Although not
commented on in the discovery paper by \citet{Stock2010},
there is a relatively bright and broad 
interstellar band of
diffuse background or foreground H$\alpha$ emission that crosses over a large portion of the
nebula's central and southern sections (see upper panel in Fig.\ 2). This emission
complicates obtaining a clear picture of the
nebula's full structure and prevents easy comparison of the nebula's
H$\alpha$ vs.\ [\ion{O}{3}] emission features
and extent. (Note: This emission 
was noted by \citet{Marston1994a} and may have prevented 
them from noticing WR~8's faint nebula.)

\begin{figure*}[ht]
\begin{center}
\includegraphics[angle=0,width=16.0cm]{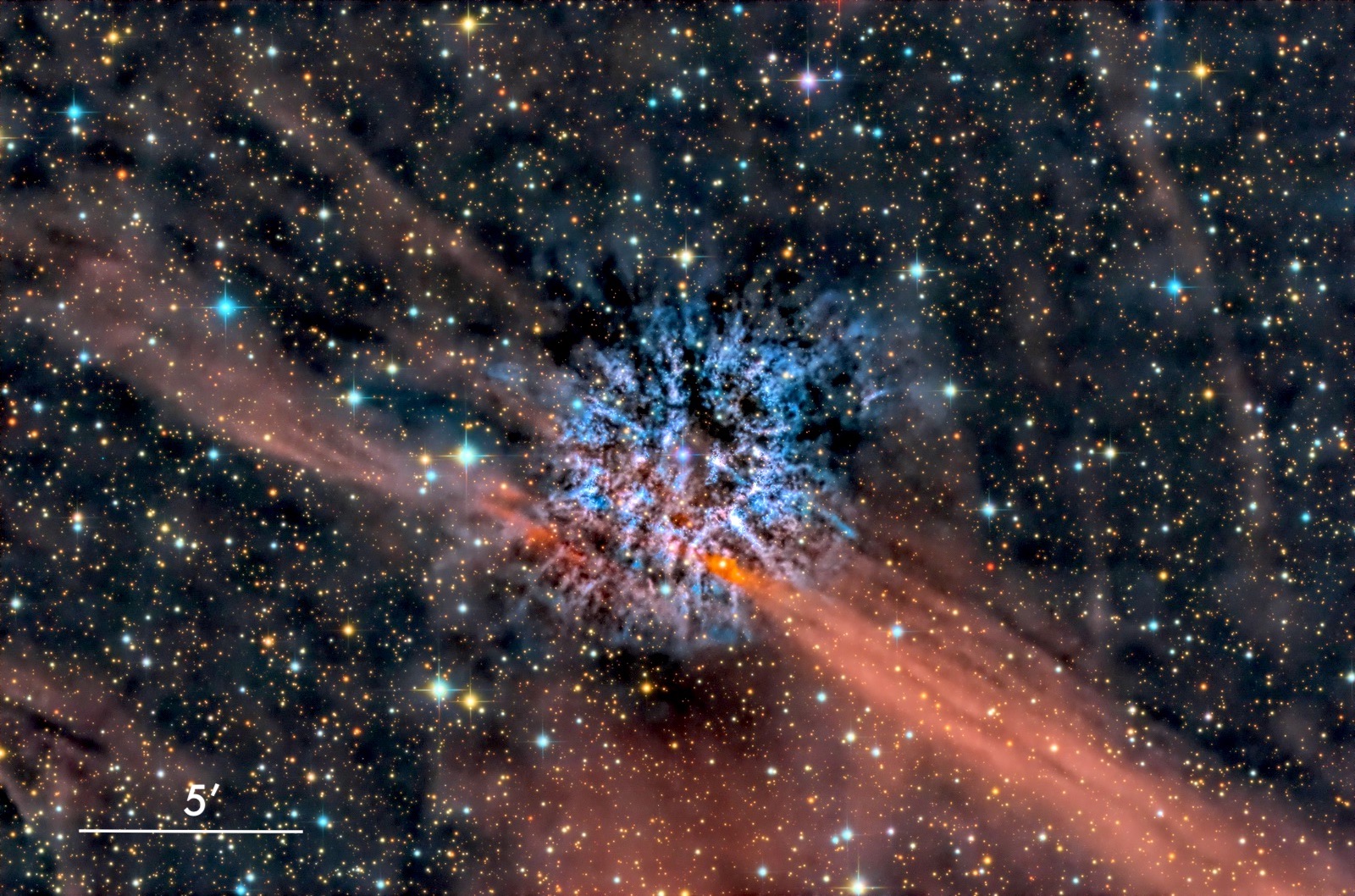}
\caption{Wide FOV color image of WR 8 composited of H$\alpha$ (red), 
[O~III] (blue) and broadband filter images.
\label{Fig3}
} 
\end{center}
\end{figure*}

\begin{figure*}
\begin{center}
\includegraphics[angle=0,width=7.8cm]{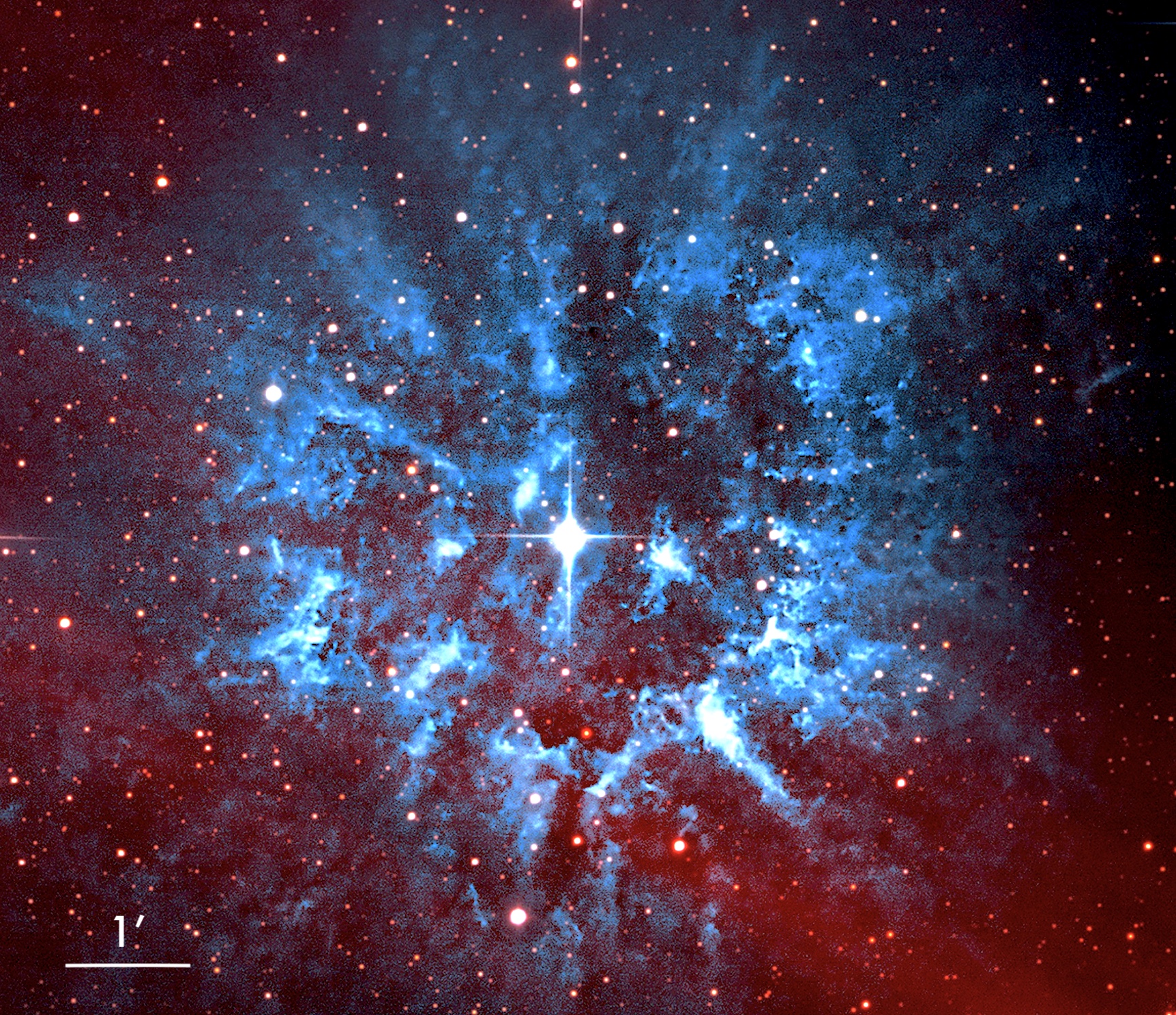}
\includegraphics[angle=0,width=7.9cm]{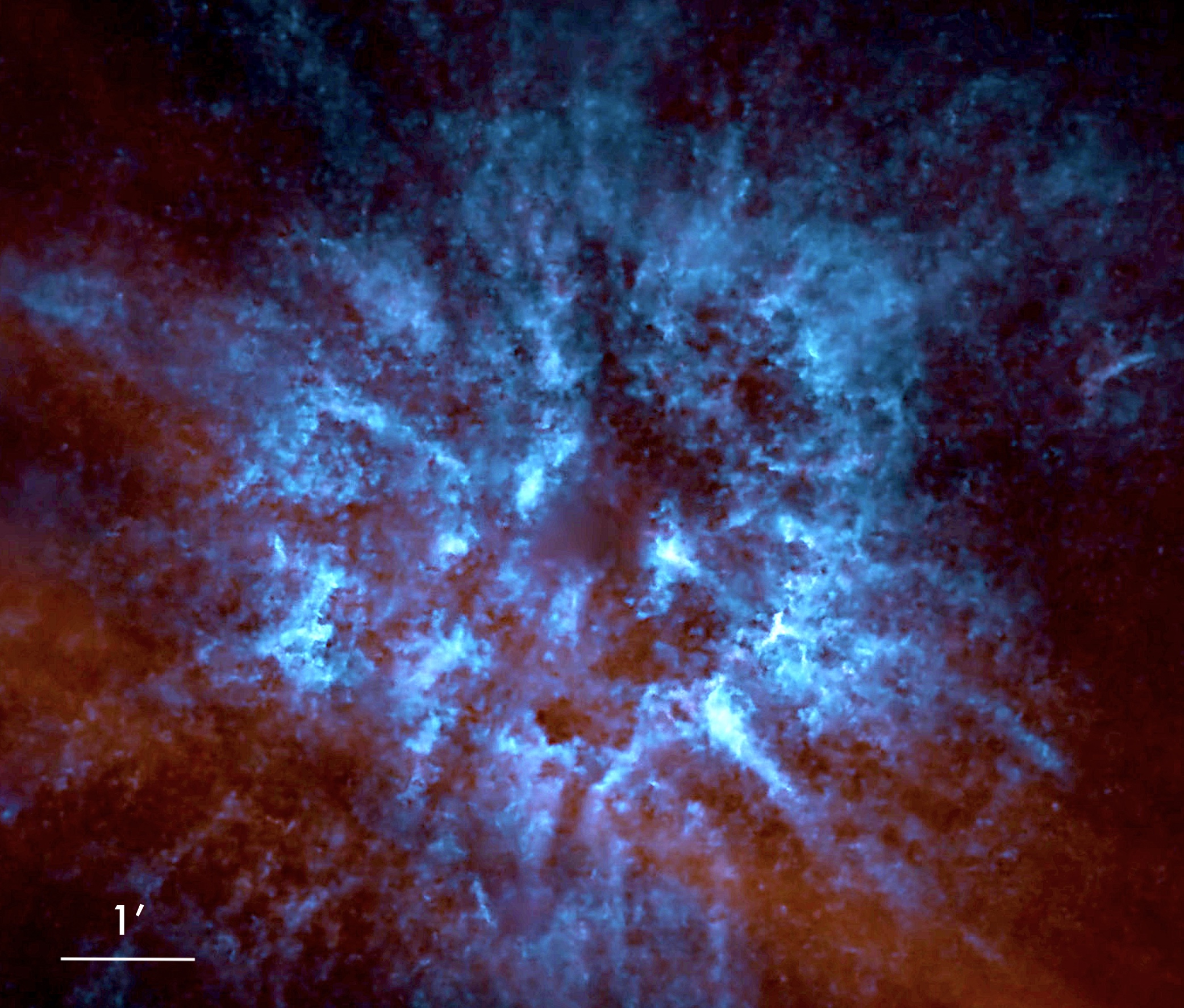}
\caption{Two independent color composite 
H$\alpha$ and [\ion{O}{3}] images of WR 8's nebula with and without stars.
\label{Fig2}
} 
\end{center}
\end{figure*}

\begin{figure*}[ht!]
\begin{center}
\includegraphics[angle=0,width=17.1cm]{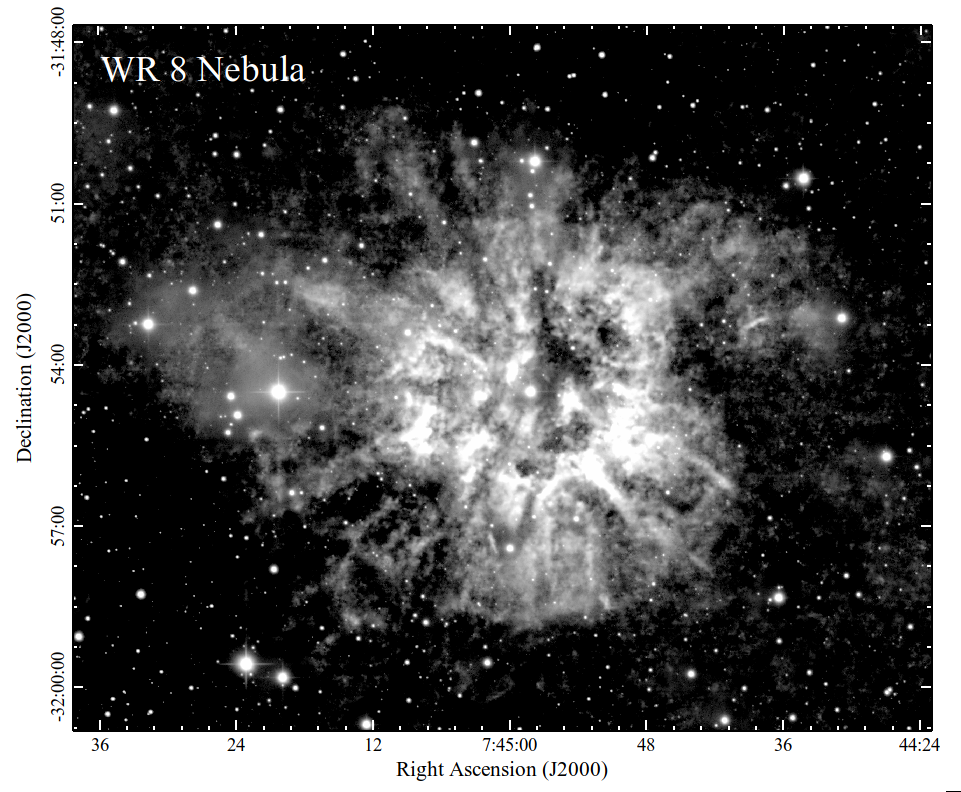}
\caption{A stretched positive [\ion{O}{3}] image
showing extended emission around WR~8's brighter inner nebula.
\label{Fig2}
} 
\end{center}
\end{figure*}

Due to this overlapping
H$\alpha$ emission, in the image 
presentations that follow, we
emphasize the nebula's [\ion{O}{3}] emission structure
over that seen in H$\alpha$ emission.
In addition, because longer 
exposures were taken
in the [\ion{O}{3}] filter in all three of 
our image sets, 
this has led us to 
construct color composites suppressing
H$\alpha$ relative to [\ion{O}{3}] emissions. In these color composites, H$\alpha$ emission is shown as red while [\ion{O}{3}] emission is blue.
The result is the WR~8
nebula is shown with a strong blue color when, in fact, the nebula's [\ion{O}{3}] is actually 
fainter than that of its H$\alpha$ emission. Indeed,
\citet{Stock2011} reported spectra that showed the nebula's [\ion{O}{3}] emission much fainter than that of H$\alpha$, with observed and 
extinction corrected F([\ion{O}{3}] $\lambda$5007)/F(H$\alpha$) 
and  I([\ion{O}{3}] $\lambda$5007)/I(H$\alpha$)  values of 0.22 and 0.79, respectively.

WR~8's nebulosity shown in 
Fig.\ 2 images may give the impression the WR~8 nebula is relatively bright when, in fact, it is quite faint. Based on its weak detection on SHS images at or near the estimated detection limit of that H$\alpha$ survey, specifically $\leq 5$ Rayleighs\footnote{Rayleigh = $3.715 \times 10^{-14}$ $\lambda^{-1}$ erg cm$^{-2}$ s$^{-1}$ arcsec$^{-2}$}, we estimate the nebula's H$\alpha$ to be roughly some 5 to 10 Rayleighs, corresponding to a H$\alpha$ flux of  $\simeq$ $2.5$ $-$  $5 \times 10^{-17}$ erg cm$^{-2}$ s$^{-1}$ arcsec$^{-2}$.

A wider view of the WR~8 nebula and interstellar emission along the line-of-sight is shown in Figure 3.
This color H$\alpha$, [\ion{O}{3}], and B,V,R filter composite of 0.6~m telescope images more clearly shows the broad interstellar band of H$\alpha$ emission along with the WR~8's nebula. The deeper [\ion{O}{3}] images combined with the desire to 
present the nebula's structure without confusion of the overlapping band of interstellar H$\alpha$ emission leads to a color composite image showing a seemingly bright [\ion{O}{3}] emission nebula.

\begin{figure*}[ht]
\begin{center}
\includegraphics[angle=0,width=8.5cm]{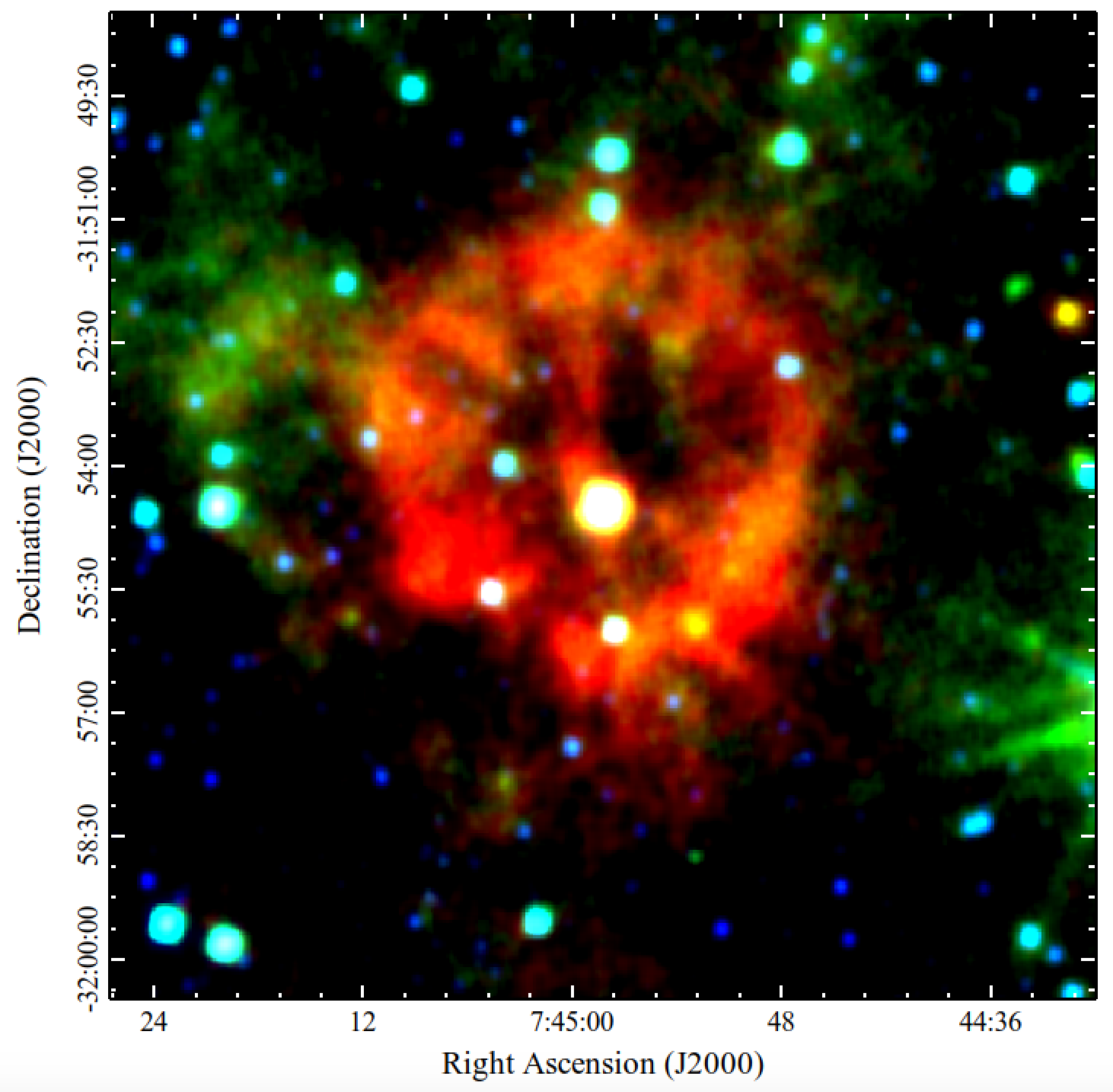}
\includegraphics[angle=0,width=8.5cm]{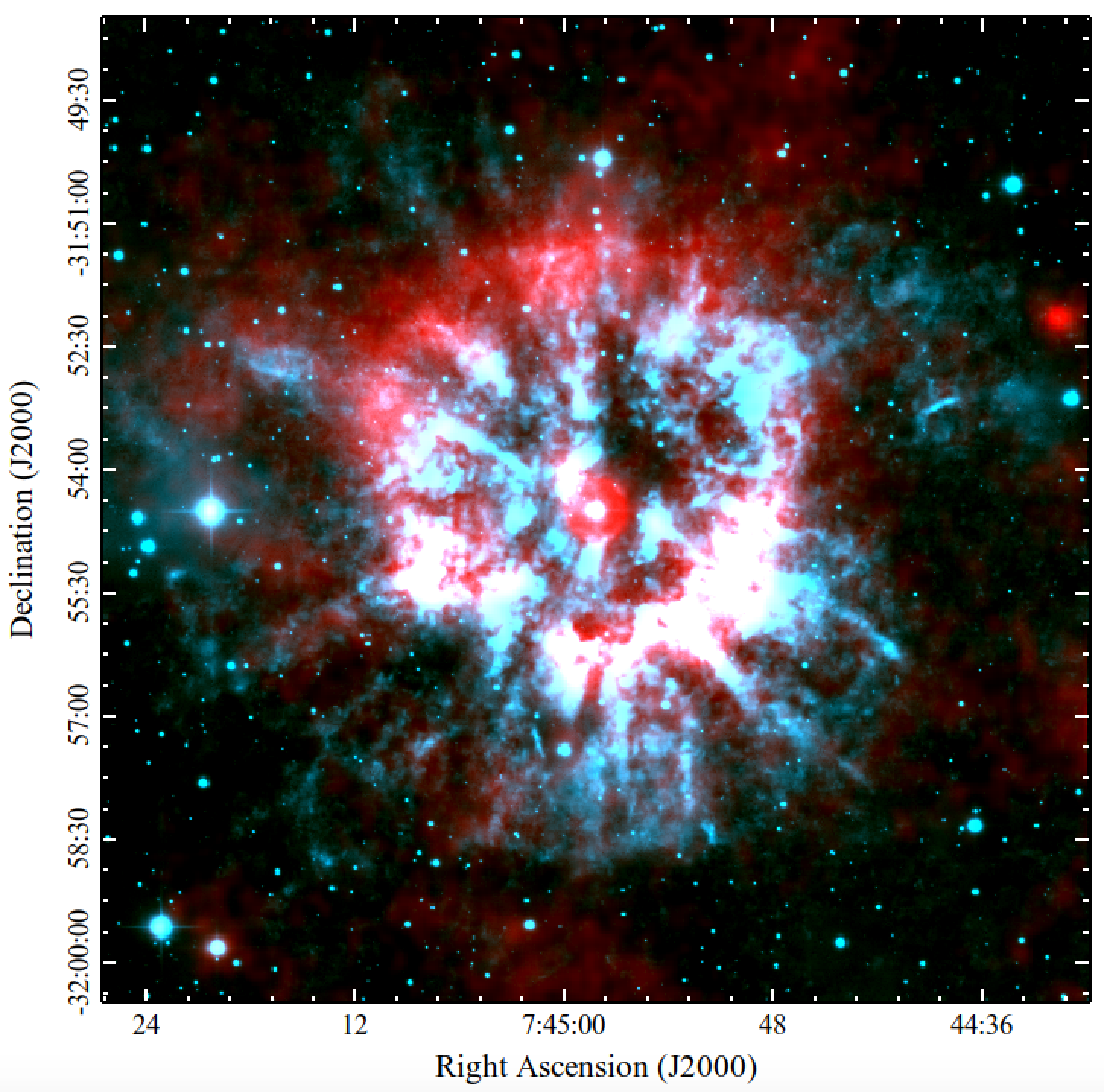}
\caption{Color composites of WR~8's nebula. Left Panel: WISE 
4.6 $\mu$m (blue), 12 $\mu$m (green), and 22 $\mu$m (red) images. \\ Right Panel: WISE 22 $\mu$m (red) 
and [\ion{O}{3}] $\lambda$5007 (blue) images.
\label{Fig2}
} 
\end{center}
\end{figure*}

The fine-scale structure of WR~8's nebula is better seen in the 
enlargements of Figure
4 where we show two independent color composite images. The left panel shows the nebula at the best resolution of our 0.7~m telescope data, while the right panel shows more of the nebula's fainter emission. Stars have been
removed from the deeper right panel image using computer software in order to make the nebula's structure more readily visible. As noted above, the nebula's H$\alpha$ emission has been 
artificially suppressed compared to its [\ion{O}{3}] emission, resulting in
overlapping  diffuse band of interstellar H$\alpha$
emission  appearing only weakly
in both panels.

Broadly, the nebula appears as a 
broken ring of numerous emission clumps the center of which 
is not exactly centered on the WR star, HD 62910.
Extending out from this emission ring are several emission 
streaks and broad fans of emission extending 
some $1' -2'$ in length which fade 
with increasing radial distance. 
The nebula's brightest features are
seen in the west-central and southwest areas 
along with some head-tail morphologies
suggestive of wind sculptured clouds of stellar ejecta.

The appearance of the WR~8 nebula resembles 
that of the much brighter ejecta rich nebula
RCW~58. Although  having far fewer in number, it exhibits some of RCW~58 radial aligned streaks. However, unlike that seen for WR~8, RCW~58's nebula
looks very different in H$\alpha$ where it is highly clumpy
compared to its [\ion{O}{3}] emission where the nebula much
more diffuse with fewer clumpy features.

\begin{deluxetable*}{lccccccl}
\tablecolumns{8}
\tablecaption{Comparison of Some Ejecta Rich Wolf-Rayet Nebulae \label{Tab2}}
\tablewidth{0pt}
\tablehead{\colhead{WR } & \colhead{Nebula} & \colhead{Wolf-Rayet} & \colhead{Spectral} & \colhead{Nebula Diameter} 
& \colhead{Distance\tablenotemark{a}} & \colhead{Physical Size} & \colhead{References} \\
         \colhead{Name}  & \colhead{Name}  & \colhead{Star}   & \colhead{Type}     & \colhead{(arcmin)}         & \colhead{(kpc)}    & \colhead{(pc)} & \colhead{} }
\startdata
  WR 6   & S308       & HD 50896    & WN4      &  40  & $1.51\pm0.09$   &  17.6     & 1, 2, 3 \\
  WR 8   &   \nodata  &  HD 62910   & WN6/WC4  &  6   & $3.52\pm0.16$   &  6.1      & 1, 4  \\ 
  WR 40  & RCW 58     &  HD 96548   & WN8      &  7 x 9 & $2.70\pm0.12 $  & 5.5 x 7.1 & 5 \\ 
  WR 71 &  \nodata    & HD 143414   & WN6      &  4   & $4.27\pm 0.39$    &  5.0      & 1, 6 \\
  WR 124 & M1-67      & BAC 209     & WN8      &  1.2 & $5.36\pm0.38$     &  1.9      &  5      \\
  WR 136 & NGC 6888   & HD 192163   & WN6      &  12 x 18 & $1.67\pm0.04$  &  5.8 x 8.7  & 1, 5, 7, 8 \\
\enddata
\tablenotetext{a}{Distances are Bayesian corrected geometric values from \citet{Bailer2021} for Gaia DR3. }
\tablenotetext{}{References: 1) \citet{Stock2010} ; 2) \citet{Esteban1992a} ; 3) \cite{Chu2003};
4) this paper;  5) \citet{Chu1983}; 6) \citet{Marston1994b}; 7) \citet{Chu1991} 
8) \citet{Esteban1992b}  }
\label{table}
\end{deluxetable*}

In contrast, WR~8's morphology is
about the same in H$\alpha$ and [\ion{O}{3}].
Also, whereas
most of RCW~58's nebula is 
concentrated in its outer regions with a relatively clear central region, WR~8's nebula has more
inner emission clouds. In this respect, it resembles
somewhat the cloudy M1-67 nebula although having far few clouds.  Finally, based upon 
their similar appearances on SHS images,
the faint and recently recognized WR~71 nebula
shares some structural features including an
incomplete and clumpy shell 
\citep{Marston1994b,Stock2010}.

Figure 5 shows a stretched version
of our deepest [\ion{O}{3}] image in  order 
to show the nebula's outermost extent.
Whereas the bright inner emission ring
spans some $5' - 6'$ in diameter,
the nebula's outermost regions have 
N-S and E-W dimensions around $9'$ 
and $13'$, respectively.
Much of the nebula's southern sections
appear to have a well  defined boundary
some 4.3$'$ from the WR star.
But that is not the situation for the other outer emission regions,
especially to the east where a large diffuse
patch of [\ion{O}{3}] emission is seen.
This extended outer and more diffuse emission is
perhaps  due to a lower mass loss rate episode in WR~8's history.

Due to the strong overlapping interstellar 
band of H$\alpha$ emission one cannot determine if the nebula's
outer regions looks the same as that seen
in [\ion{O}{3}]. 
What is clear, however, is that the nebula exhibits a somewhat similar clumpy but yet different 
appearance in the infrared from that seen in optical line emissions.

As  noted by \citet{Toala2015}, WR~8 possesses
a bright nebular ring at 22 $\mu$m.
This can be seen in  the left panel of Fig.\ 6 where we  combined WISE blue
4.6 $\mu$m, green 12 $\mu$m, and red 22 $\mu$m images into a color
composite of the WR~8 nebula.
Much like that seen in the optical, this infrared 
emission mainly due to thermal continuum emission from dust \citep{Toala2015}
shows a fairly clumpy emission structure
but in a better defined shell.
The geometric center of this shell lies 
some $40''$ north of ER~8, making the star appear noticeably
off-center.

The right hand panel of Fig.\ 6 shows
a color composite of 
our [\ion{O}{3}] $\lambda$5007 image in blue with the
WISE 22 $\mu$m image
in red. 
The clumpy structure seen in the infrared image is seen to be
associated with many of the clumpy features in our
[\ion{O}{3}] emission images,
especially along the nebula's
northern rim.
However, the gross morphology differences between WR~8's infrared nebula and its [\ion{O}{3}] emissions indicate  differences of where
WR~8's nebula's thermal dust emission lies vs.\ its ionized gas.  

Unlike the general similarity
found by \citet{Toala2015} for  H$\alpha$ and 22 $\mu$m emissions in most WR ring nebulae,
WR~8's nebula appears especially noticeably different along the
[\ion{O}{3}] 
clumps and streamers along its northern rim.
Whereas the nebula's warm dust as detected in the 22 $\mu$m image lies coincident with WR~8's bright optical
features in south,  its 22 $\mu$m  emission is mainly seen 
in the fainter outer portions of 
bright [\ion{O}{3}] northern emission clumps. 
It is this greater radial
distance of dust emission along the nebula's northern areas
that is responsible for the striking off-center
position of the
WR star from the middle of the IR shell. We further note that a
stronger intensity stretch of the 22 $\mu$m image shows fainter IR  emission surrounding much of this otherwise well defined shell, especially to the south and southwest which is coincident with some
very faint [\ion{O}{3}] emission.

The cause for the stark
coincidence difference of the nebula's dust
in and around the nebula's ionized gas clumps in the south compared to what is seen in the north is unclear. Our [\ion{O}{3}] images
show considerably more and brighter emission in the south compared to the north (see Figs.\ 2 and 4),
raising the possibility of differential dust temperatures in the north vs.\ south assuming a constant gas to dust ratio.
Greater gas densities and optical depths in gasous clumps located along the shell's southern rim might give rise to cooler dust relative to warmer dust in the less dense and less UV shielded northern regions.
However, this might not explain the weak infrared emission for some of the nebula's other optically bright features such as in its east-central region.

Finally, one can ask how does WR~8's
nebula's physical size compare to other ejecta dominated
ring nebula. In Table 1, we list the 
WR~8's nebula compared to 
several WR ring nebula rich in stellar ejecta
indicated by over-abundances of  nitrogen and helium as markers
of the CNO process of energy production in high mass stars.
One sees that this small and select group of E-type
WR nebulae have fairly similar physical diameters of around 5 to 15 pc. Although WR~6's S308 nebula is the largest of this group, WR~8 nebula's faint outer angular dimensions of $9' \times 13'$  corresponding to roughly $9 \times 14$ pc is not that far behind. We note that the  M1-67 nebula stands out from the rest as the smallest, presumably due to its younger age.

Thus it seems that WR~8's nebula as detected in our images
is not unlike those of other ejecta rich WR nebulae.
While \citet{Stock2011} did not report on the expansion velocity of WR 8's nebula.
if we choose an expansion velocity like that seen for other WR nebulae, namely of order 60 km s$^{-1}$
\citep{Smith1994, Siri1998}, then the $\simeq$3 pc radius of its bright
main shell suggests a dynamical age of around 50 kyr.


In summary, our optical line emission images show WR~8 to possess a striking and
extensive optical nebula consisting of a brighter
inner ring of emission surrounded by an
outer nebula of fainter emission best seen
in [\ion{O}{3}]. 
Overall, the WR~8 nebula is especially visually impressive 
compared to many other WR ring nebulae. 
Being one of only a few WR stars having a stellar ejecta dominated nebula and being associated with
a WR star in transition from WN to WC, 
its nebula deserves further study.
An in depth spectral
investigation of its main shell and outer halo of emission may prove helpful in improving our understanding of the brief WN to WC
transition phase. \\

We would like to thank Phil Massey 
for encouraging our imaging of WR~8, an anonymous referee for several helpful suggestions, and Yuri Beletsky
of Las Campanas Observatory for help obtaining the IMACS observations on WR~8. 
This work is part of R.A.F's Archangel III Research Program at Dartmouth. 

\facilities{Las Campanas Observatory, Observatorio El Sauce, Chile}
\software{Photoshop, PixInsight, Astropixel 
Processor, DS9 fits viewer \citet{Joye2003}, WCSTools \citet{Laycock2010}  }
 

\bibliography{casa.bib}

\begin{thebibliography}{}
\expandafter\ifx\csname natexlab\endcsname\relax\def\natexlab#1{#1}\fi
\providecommand{\url}[1]{\href{#1}{#1}}
\providecommand{\dodoi}[1]{doi:~\href{http://doi.org/#1}{\nolinkurl{#1}}}
\providecommand{\doeprint}[1]{\href{http://ascl.net/#1}{\nolinkurl{http://ascl.net/#1}}}
\providecommand{\doarXiv}[1]{\href{https://arxiv.org/abs/#1}{\nolinkurl{https://arxiv.org/abs/#1}}}

\bibitem[{{Abbott}(1982)}]{Abbott82}
{Abbott}, D.~C. 1982, \apj, 263, 723, \dodoi{10.1086/160544}

\bibitem[{{Abbott} \& {Conti}(1987)}]{Abbott1987}
{Abbott}, D.~C., \& {Conti}, P.~S. 1987, \araa, 25, 113, \dodoi{10.1146/annurev.aa.25.090187.000553}

\bibitem[{{Bailer-Jones} {et~al.}(2021){Bailer-Jones}, {Rybizki}, {Fouesneau}, {Demleitner}, \& {Andrae}}]{Bailer2021}
{Bailer-Jones}, C.~A.~L., {Rybizki}, J., {Fouesneau}, M., {Demleitner}, M., \& {Andrae}, R. 2021, \aj, 161, 147, \dodoi{10.3847/1538-3881/abd806}

\bibitem[{{Chu}(1981)}]{Chu1981}
{Chu}, Y.~H. 1981, \apj, 249, 195, \dodoi{10.1086/159275}

\bibitem[{{Chu}(1991)}]{Chu1991}
{Chu}, Y.~H. 1991, in IAU Symposium, Vol. 143, Wolf-Rayet Stars and Interrelations with Other Massive Stars in Galaxies, ed. K.~A. {van der Hucht} \& B.~{Hidayat}, 349

\bibitem[{{Chu}(2016)}]{Chu2016}
{Chu}, Y.-H. 2016, in Journal of Physics Conference Series, Vol. 728, Journal of Physics Conference Series (IOP), 032007, \dodoi{10.1088/1742-6596/728/3/032007}

\bibitem[{{Chu} {et~al.}(2003){Chu}, {Guerrero}, {Gruendl}, {Garc{\'\i}a-Segura}, \& {Wendker}}]{Chu2003}
{Chu}, Y.-H., {Guerrero}, M.~A., {Gruendl}, R.~A., {Garc{\'\i}a-Segura}, G., \& {Wendker}, H.~J. 2003, \apj, 599, 1189, \dodoi{10.1086/379607}

\bibitem[{{Chu} {et~al.}(1983){Chu}, {Treffers}, \& {Kwitter}}]{Chu1983}
{Chu}, Y.~H., {Treffers}, R.~R., \& {Kwitter}, K.~B. 1983, \apjs, 53, 937, \dodoi{10.1086/190914}

\bibitem[{{Conti}(2000)}]{Conti2000}
{Conti}, P.~S. 2000, \pasp, 112, 1413, \dodoi{10.1086/317693}

\bibitem[{{Conti} \& {Massey}(1989)}]{Conti_Massey1989}
{Conti}, P.~S., \& {Massey}, P. 1989, \apj, 337, 251, \dodoi{10.1086/167101}

\bibitem[{{Crowther}(2007)}]{Crowther2007}
{Crowther}, P.~A. 2007, \araa, 45, 177, \dodoi{10.1146/annurev.astro.45.051806.110615}

\bibitem[{{Crowther}(2008)}]{Crowther2008}
{Crowther}, P.~A. 2008, in IAU Symposium, Vol. 250, Massive Stars as Cosmic Engines, ed. F.~{Bresolin}, P.~A. {Crowther}, \& J.~{Puls}, 47--62, \dodoi{10.1017/S1743921308020334}

\bibitem[{{Deshmukh} {et~al.}(2024){Deshmukh}, {Sana}, {M{\'e}rand}, {Bordier}, {Langer}, {Bodensteiner}, {Dsilva}, {Frost}, {Gosset}, {Le Bouquin}, {Lefever}, {Mahy}, {Patrick}, {Reggiani}, {Sander}, {Shenar}, {Tramper}, {Villase{\~n}or}, \& {Waisberg}}]{Deshmukh2024}
{Deshmukh}, K., {Sana}, H., {M{\'e}rand}, A., {et~al.} 2024, \aap, 692, A109, \dodoi{10.1051/0004-6361/202452352}

\bibitem[{{Dressler} {et~al.}(2011){Dressler}, {Bigelow}, {Hare}, {Sutin}, {Thompson}, {Burley}, {Epps}, {Oemler}, {Bagish}, {Birk}, {Clardy}, {Gunnels}, {Kelson}, {Shectman}, \& {Osip}}]{Dressler2011}
{Dressler}, A., {Bigelow}, B., {Hare}, T., {et~al.} 2011, \pasp, 123, 288, \dodoi{10.1086/658908}

\bibitem[{{Esteban} {et~al.}(2016){Esteban}, {Mesa-Delgado}, {Morisset}, \& {Garc{\'\i}a-Rojas}}]{Esteban2016}
{Esteban}, C., {Mesa-Delgado}, A., {Morisset}, C., \& {Garc{\'\i}a-Rojas}, J. 2016, \mnras, 460, 4038, \dodoi{10.1093/mnras/stw1243}

\bibitem[{{Esteban} \& {Vilchez}(1992)}]{Esteban1992b}
{Esteban}, C., \& {Vilchez}, J.~M. 1992, \apj, 390, 536, \dodoi{10.1086/171303}

\bibitem[{{Esteban} {et~al.}(1992){Esteban}, {Vilchez}, {Smith}, \& {Clegg}}]{Esteban1992a}
{Esteban}, C., {Vilchez}, J.~M., {Smith}, L.~J., \& {Clegg}, R.~E.~S. 1992, \aap, 259, 629

\bibitem[{{Hamann} {et~al.}(2019){Hamann}, {Gr{\"a}fener}, {Liermann}, {Hainich}, {Sander}, {Shenar}, {Ramachandran}, {Todt}, \& {Oskinova}}]{Hamann2019}
{Hamann}, W.~R., {Gr{\"a}fener}, G., {Liermann}, A., {et~al.} 2019, \aap, 625, A57, \dodoi{10.1051/0004-6361/201834850}

\bibitem[{{Heckathorn} {et~al.}(1982){Heckathorn}, {Bruhweiler}, \& {Gull}}]{Heckathorn1982}
{Heckathorn}, J.~N., {Bruhweiler}, F.~C., \& {Gull}, T.~R. 1982, in IAU Symposium, Vol.~99, Wolf-Rayet Stars: Observations, Physics, Evolution, ed. C.~W.~H. {De Loore} \& A.~J. {Willis}, 463

\bibitem[{{Hillier} {et~al.}(2021){Hillier}, {Aadland}, {Massey}, \& {Morrell}}]{Hillier2021}
{Hillier}, D.~J., {Aadland}, E., {Massey}, P., \& {Morrell}, N. 2021, \mnras, 503, 2726, \dodoi{10.1093/mnras/stab580}

\bibitem[{{Johnson} \& {Hogg}(1965)}]{Johnson1965}
{Johnson}, H.~M., \& {Hogg}, D.~E. 1965, \apj, 142, 1033, \dodoi{10.1086/148373}

\bibitem[{{Joye} \& {Mandel}(2003)}]{Joye2003}
{Joye}, W.~A., \& {Mandel}, E. 2003, in Astronomical Society of the Pacific Conference Series, Vol. 295, Astronomical Data Analysis Software and Systems XII, ed. H.~E. {Payne}, R.~I. {Jedrzejewski}, \& R.~N. {Hook}, 489

\bibitem[{{Kwitter}(1984)}]{Kwitter1984}
{Kwitter}, K.~B. 1984, \apj, 287, 840, \dodoi{10.1086/162742}

\bibitem[{{Laycock} {et~al.}(2010){Laycock}, {Tang}, {Grindlay}, {Los}, {Simcoe}, \& {Mink}}]{Laycock2010}
{Laycock}, S., {Tang}, S., {Grindlay}, J., {et~al.} 2010, \aj, 140, 1062, \dodoi{10.1088/0004-6256/140/4/1062}

\bibitem[{{Maeder} \& {Conti}(1994)}]{Maeder1994}
{Maeder}, A., \& {Conti}, P.~S. 1994, \araa, 32, 227, \dodoi{10.1146/annurev.astro.32.1.227}

\bibitem[{{Marchenko} {et~al.}(1998){Marchenko}, {Moffat}, {Eversberg}, {Morel}, {Hill}, {Tovmassian}, \& {Seggewiss}}]{Marchenko1998}
{Marchenko}, S.~V., {Moffat}, A.~F.~J., {Eversberg}, T., {et~al.} 1998, \mnras, 294, 642, \dodoi{10.1111/j.1365-8711.1998.01174.x10.1046/j.1365-8711.1998.01174.x}

\bibitem[{{Marston} {et~al.}(1994{\natexlab{a}}){Marston}, {Chu}, \& {Garcia-Segura}}]{Marston1994a}
{Marston}, A.~P., {Chu}, Y.~H., \& {Garcia-Segura}, G. 1994{\natexlab{a}}, \apjs, 93, 229, \dodoi{10.1086/192053}

\bibitem[{{Marston} {et~al.}(1994{\natexlab{b}}){Marston}, {Yocum}, {Garcia-Segura}, \& {Chu}}]{Marston1994b}
{Marston}, A.~P., {Yocum}, D.~R., {Garcia-Segura}, G., \& {Chu}, Y.~H. 1994{\natexlab{b}}, \apjs, 95, 151, \dodoi{10.1086/192097}

\bibitem[{{Niemela}(1991)}]{Niemela1991}
{Niemela}, V.~S. 1991, in IAU Symposium, Vol. 143, Wolf-Rayet Stars and Interrelations with Other Massive Stars in Galaxies, ed. K.~A. {van der Hucht} \& B.~{Hidayat}, 201

\bibitem[{{Oey}(1999)}]{Oey1999}
{Oey}, M.~S. 1999, in IAU Symposium, Vol. 193, Wolf-Rayet Phenomena in Massive Stars and Starburst Galaxies, ed. K.~A. {van der Hucht}, G.~{Koenigsberger}, \& P.~R.~J. {Eenens}, 627

\bibitem[{{Parker} {et~al.}(2005){Parker}, {Phillipps}, {Pierce}, {Hartley}, {Hambly}, {Read}, {MacGillivray}, {Tritton}, {Cass}, {Cannon}, {Cohen}, {Drew}, {Frew}, {Hopewell}, {Mader}, {Malin}, {Masheder}, {Morgan}, {Morris}, {Russeil}, {Russell}, \& {Walker}}]{Parker2005}
{Parker}, Q.~A., {Phillipps}, S., {Pierce}, M.~J., {et~al.} 2005, \mnras, 362, 689, \dodoi{10.1111/j.1365-2966.2005.09350.x}

\bibitem[{{Sander} {et~al.}(2012){Sander}, {Hamann}, \& {Todt}}]{Sander2012}
{Sander}, A., {Hamann}, W.~R., \& {Todt}, H. 2012, \aap, 540, A144, \dodoi{10.1051/0004-6361/201117830}

\bibitem[{{Sander} {et~al.}(2022){Sander}, {Vink}, {Higgins}, {Shenar}, {Hamann}, \& {Todt}}]{Sander2022}
{Sander}, A. A.~C., {Vink}, J.~S., {Higgins}, E.~R., {et~al.} 2022, in IAU Symposium, Vol. 366, The Origin of Outflows in Evolved Stars, ed. L.~{Decin}, A.~{Zijlstra}, \& C.~{Gielen}, 21--26, \dodoi{10.1017/S1743921322000400}

\bibitem[{{Sirianni} {et~al.}(1998){Sirianni}, {Nota}, {Pasquali}, \& {Clampin}}]{Siri1998}
{Sirianni}, M., {Nota}, A., {Pasquali}, A., \& {Clampin}, M. 1998, \aap, 335, 1029

\bibitem[{{Smith}(1994)}]{Smith1994}
{Smith}, L.~J. 1994, \apss, 216, 291, \dodoi{10.1007/BF00982507}

\bibitem[{{Stock} \& {Barlow}(2010)}]{Stock2010}
{Stock}, D.~J., \& {Barlow}, M.~J. 2010, \mnras, 409, 1429, \dodoi{10.1111/j.1365-2966.2010.17124.x}

\bibitem[{{Stock} \& {Barlow}(2014)}]{Stock2014}
---. 2014, \mnras, 441, 3065, \dodoi{10.1093/mnras/stu724}

\bibitem[{{Stock} {et~al.}(2011){Stock}, {Barlow}, \& {Wesson}}]{Stock2011}
{Stock}, D.~J., {Barlow}, M.~J., \& {Wesson}, R. 2011, \mnras, 418, 2532, \dodoi{10.1111/j.1365-2966.2011.19643.x}

\bibitem[{{Toal{\'a}} {et~al.}(2015){Toal{\'a}}, {Guerrero}, {Ramos-Larios}, \& {Guzm{\'a}n}}]{Toala2015}
{Toal{\'a}}, J.~A., {Guerrero}, M.~A., {Ramos-Larios}, G., \& {Guzm{\'a}n}, V. 2015, \aap, 578, A66, \dodoi{10.1051/0004-6361/201525706}

\bibitem[{{van der Hucht}(2001)}]{vanderHucht2001}
{van der Hucht}, K.~A. 2001, \nar, 45, 135, \dodoi{10.1016/S1387-6473(00)00112-3}

\bibitem[{{van der Hucht} {et~al.}(1981){van der Hucht}, {Conti}, {Lundstrom}, \& {Stenholm}}]{vander1981}
{van der Hucht}, K.~A., {Conti}, P.~S., {Lundstrom}, I., \& {Stenholm}, B. 1981, \ssr, 28, 227, \dodoi{10.1007/BF00173260}

\bibitem[{{Wachter} {et~al.}(2011){Wachter}, {Cohen}, \& {Leisawitz}}]{Wachter2011}
{Wachter}, S., {Cohen}, M., \& {Leisawitz}, D. 2011, in American Astronomical Society Meeting Abstracts, Vol. 217, 333.10

\bibitem[{{Willis} \& {Stickland}(1990)}]{Willis1990}
{Willis}, A.~J., \& {Stickland}, D.~J. 1990, \aap, 232, 89

\end{thebibliography}

\end{document}